\documentclass[preprint,12pt]{elsarticle}

\usepackage{multicol}

\usepackage{graphicx}
\usepackage{amssymb}

\usepackage{lineno}

\begin{document}

\begin{frontmatter}


\title{Quantitative assessment of changes in cellular morphology at photodynamic treatment \textit{in vitro} by means of digital holographic microscopy
}



\author[label1]{A.V. Belashov}
\author[label1]{A.A. Zhikhoreva}
\author[label2]{T.N. Belyaeva}
\author[label2,label3]{E.S. Kornilova}
\author[label2]{A.V. Salova}
\author[label1]{I.V. Semenova, \corref{cor1}}
\author[label1]{O.S. Vasyutinskii}

\address[label1]{Ioffe Institute; 26, Polytekhnicheskaya, St.Petersburg, 194021, Russia}
\address[label2]{Institute of Cytology of RAS; 4, Tikhoretsky pr., St.Petersburg, 194064, Russia}
\address[label3]{Peter the Great St.Petersburg Polytechnic University; 29, Polytekhnicheskaya, St.Petersburg, 195251, Russia}

\cortext[cor1]{irina.semenova@mail.ioffe.ru}

\begin{abstract}
Changes in morphological characteristics of cells from two cultured cancer cell lines, HeLa and A549, induced by photodynamic treatment with Radachlorin photosensitizer have been monitored using digital holographic microscopy. The observed dose-dependent post-treatment dynamics of phase shift variations demonstrated several scenarios of cell death. In particular the phase shift increase at low doses can be associated with apoptosis while its decrease at high doses can be associated with necrosis.
Two cell types were shown to be differently responsive to treatment at the same doses. Although the sequence of death scenarios with increasing irradiation dose was demonstrated to be the same, each specific scenario was realized at substantially different doses.
Results obtained by holographic microscopy were confirmed by confocal fluorescence microscopy with the commonly used test assay.
\end{abstract}

\begin{keyword}
quantitative phase imaging \sep digital holographic microscopy \sep cell death pathways \sep photodynamic treatment \sep HeLa \sep A549


\end{keyword}

\end{frontmatter}


\section{Introduction}
\label{S:1}

A persistent increase of cancer incidences  and high recurrence rate are key problems in modern oncology requiring improvement of existing treatment modalities and development of novel approaches for early diagnostics and therapy. One of highly promising modalities is photodynamic therapy (PDT) which was already successfully applied for treatment of various malignant and benign pathologies, skin lesions, macular degeneration, microbial infections etc. \cite{Wilson2008, Tayyaba2004, Letai2017}.  PDT employs specific physical-chemical properties of molecular photosensitizers (PSs) which are known to be selectively accumulated in pathological tissues with enhanced metabolism. Excitation of PS molecules by light within an absorption band leads to the formation of reactive oxygen species (ROS) which cause cell death, tumor resorption and ablastics of the lesion.

The anti-tumor effect of PDT is provided by the three interconnected processes: direct death of tumor cells, vascular disruption and activation of immune response \cite{Firczuk2011PDT}. Although the integral effect of PDT is known relatively well, contributions and mechanisms of occurring processes are still far from being sufficiently understood \cite{Brackett2011PDT,Hamblin2011}. Moreover it was presumed that the contribution of each of these processes to the tumor response to PDT may vary depending upon the PS type and PDT protocol  parameters, irradiation dose and duration in particular.
The analysis of individual cells' response to photodynamic (PD) treatment at various conditions is thus an essential aim for the study of PS efficacy as a  prerequisite for evaluation of optimal treatment doses.

A considerable progress has recently been achieved in characterization of cell death mechanisms and pathways. The commonly used classification of cell death through apoptosis and necrosis has been added by more subtle distinction, and several novel mechanisms of cell death have been introduced, see e.g. \cite{Hamblin2011,Su_2016,Zhivotovsky_2017}. The Nomenclature Committee on Cell Death has recently released definitions for four typical (apoptosis, necrosis, autophagic cell death and cornification) and eight atypical mechanisms of cell death \cite{Classification_2009}.  It is
worth noting that a cell may even switch back and forth between different death pathways \cite{Manda2009}.

From the viewpoint of present-day research on cellular response to PD treatment the reliable distinction between apoptotic and necrotic pathways along with determination of corresponding treatment doses is still  highly desirable. While necrosis is considered as a quick, violent and unprogrammed cell death caused by excessive chemical or physical impact, apoptosis is a cell suicide developing in accordance to a specific program. Necrosis is characterized by cytoplasm swelling, organelles destruction and plasma membrane disintegration, which results in the efflux of intracellular contents, causing \textit{in vivo} inflammation. On the contrary, apoptosis involves cell shrinkage, more tight packing of cytoplasm and organelles, that is followed by extensive plasma membrane blebbing and formation of separate apoptotic bodies which are \textit{in vivo} phagocytosed by macrophages or adjacent normal cells \cite{Elmore2007, Robertson_2009}. Determination of treatment doses providing cell death mainly through apoptosis is highly desirable for medical implementations since it is less harmful for patients and does not cause further inflammatory reaction.

  The distinction between apoptotic and necrotic behavior of cells is usually made by assessment of cell membrane integrity using specific standard test assays and further analysis by confocal fluorescent microscopy. Other approaches are based on determination of morphological changes of cells by means of flow cytometry or light and transmission electron microscopy (TEM). Note that until quite recently TEM was considered a "gold standard" to confirm apoptosis \cite{Elmore2007}. However most of these methods (TEM microscopy, flow cytometry) do not allow for monitoring cellular changes in dynamics, they rather provide information on cell condition in a certain time period. Besides that fluorescence-based techniques necessarily require specific fluorescent probes which can alter cellular characteristics.
Optical techniques operating with phase variations of the recording radiation passing through the object are nondestructive and allow for monitoring cellular changes in dynamics. The oldest method from this group is phase contrast microscopy, widely applied in cellular research. This methodology allows for obtaining cell images without any staining with much higher image contrast than  regular light microscopy. However quantitative estimations of cellular morphology are still problematic with this technique. Much more informative are techniques of quantitative phase imaging, digital holography in particular, which were already widely demonstrated to be very advantageous in research of various processes at the cellular level (see e.g. \cite{lee2013quantitative,park2008refractive,Rybnikov'13,PopescuChapter2008,Kuhn2013}). In \cite{OL2016} we presented our first results on determination of cellular morphology at PD treatment by means of digital holographic microscopy. The necrotic pathway of cell death was investigated in two types of cell cultures: cancer and stem.

In this paper we present results on comprehensive monitoring of variations in morphological characteristics of the two widely used cultured cancer cell lines in the course and after PD treatment with chlorin PS at various irradiation doses.  The post-treatment variations of cellular morphology were monitored by means of digital holographic microscopy.
High-precision measurements of phase shift gained by probe radiation in targeted cells demonstrate changes of their volume in the course and after PD treatment.
The phase shift dynamics has been analyzed as function of treatment parameters.
The observed dose-dependent post-treatment dynamics of phase shift variations demonstrated several scenarios of cell death.
Experiments performed by digital microscopy were assisted by observations of the same cells by far-field microscopy. The cell membrane integrity was examined by the  commonly used Acridine orange and Etidium bromide (AO/EB) test assay with observation of its fluorescence using confocal fluorescence microscopy.

\section{Experimental approach}
\subsection{Specimen preparation}

Investigations of living cells' response to photodynamic treatment  was performed on two cultured cell lines: human cervix epidermoid carcinoma HeLa cells and human alveolar basal epithelial adenocarcinoma A549 cells (both from the Russian Cell Culture Collection, Institute of Cytology RAS, St. Petersburg, Russia). Cells were cultivated in the Dulbecco's modified Eagle medium (DMEM) supplemented with 10\% fetal bovine serum and 1\% penicillin-streptomycin at 37$^o$C in 5\% CO$_2$ atmosphere. In 48 h after seeding on Petri dishes Radachlorin PS (RadaPharma, Russia) was added to the culture medium at final concentration of 5 $\mu$g/ml. Cells were incubated in this solution for 4 hours, then the  medium was replaced by the one without PS. As shown in \cite{OL2016,Biswas2014} Radachlorin penetrates through the cellular membrane and accumulates mainly in mitochondria, lysosomes and endoplasmic reticulum. Basic photophysical properties of Radachlorin in aqueous solution were studied in our recent works  \cite{SAA_time-res_2017,CPL_bleaching_2016}, where the efficient generation of singlet oxygen was demonstrated.

PS-loaded cells were irradiated by a diode laser operating at 660 nm, close to the maximum of the Q absorption band of the PS. The laser beam fluence rate was within the range of 6-130 mW/cm$^2$.
To provide nondestructive monitoring of cell parameters the holograms recording was carried out using a low-power CW HeNe laser operating at 633 nm, outside the PS absorption bands. The recording radiation fluence rate was maintained at about 50 $\mu W/cm ^2$.

\subsection{Digital holograms recording and processing}

Changes of cellular morphology resulted from PD treatment were monitored by means of inverted  digital holographic microscope in the off-axis Mach-Zehnder layout (see \cite{OL2016} for details). The 20x microscope objective and collimating lens in the object channel provided spatial resolution of about 0.8 $\mu m$. Relatively low magnification allowed us to observe several cells at each phase image.
Automatic scanning of the sample was performed using a two-coordinate motorized stage (Standa); holograms were recorded by a Videoscan-205 CCD camera (Videoscan) controlled by a software designed in the LabView 8.5 development system. Iterative monitoring of several dozens of specimen areas was carried out every five minutes during 1.5 hours after irradiation.
To increase the quality of obtained phase images several digital holograms were recorded at each position of the motorized stage and the one with highest contrast was used for reconstruction.
Hologram quality was assessed by calculation of the total intensity in the 1-st diffraction order. This allows to obtain high-quality phase images even in conditions of minor vibrations, caused by motorized stage movement and other sources. Each set of digital holograms recorded at different specimen areas also contained image of the area without cells. This phase distribution, also referred to as "background" phase image, was subtracted from each reconstructed phase distribution for compensation of optical system aberrations. Reconstruction of the recorded digital holograms was performed by means of the least square estimation algorithm \cite{liebling2004,oe2014} based on the assumption of slowly varying phase and amplitude distributions of the object wave. For biological objects, living cells in particular, this hypothesis is almost always fulfilled due to smooth shapes and high transparency of cells. Additional filtration of the obtained phase images was performed for images with  high values of shot or coherent noise. Since in this research high spatial resolution was not required, smoothing of phase images was done by means of the sin-cos algorithm \cite{aebischer1999simple}.
Afterwards, the obtained phase distributions were unwrapped using Goldstein algorithm \cite{goldstein1988satellite}. So far as phase shift is a relative value, each unwrapped phase image was normalized to make sure that image areas  containing no cells induce an approximately zero phase shift. At the final step of phase image processing cells segmentation was performed.

\subsection{Recording of cellular morphology}

Cultured cells are characterized by heterogeneity due to several reasons, such as asynchronous cell cycle and difference in individual cell shapes. These features result in significant variations of initial phase shift introduced by individual cells and of their response to external stimuli,  such as e.g. PD treatment.
Therefore several dozens of cells were monitored for obtaining a statistically significant mean value for each cell sample. Statistical analysis of the data obtained provided robust information on the typical cellular response to PD treatment. The measurement accuracy was improved by monitoring morphological changes of the same living cells at each time point during the experiment. A significant decrease of potential errors due to diversity of initial cellular parameters was thus achieved.

The confocal fluorescence microscope Leica TCS SP5 was utilized for additional monitoring of cells condition after PD treatment and for control of the cell membranes integrity.

\section{Experimental results}

\begin{figure}[htbp]
\centering
\includegraphics[width=13.5cm]{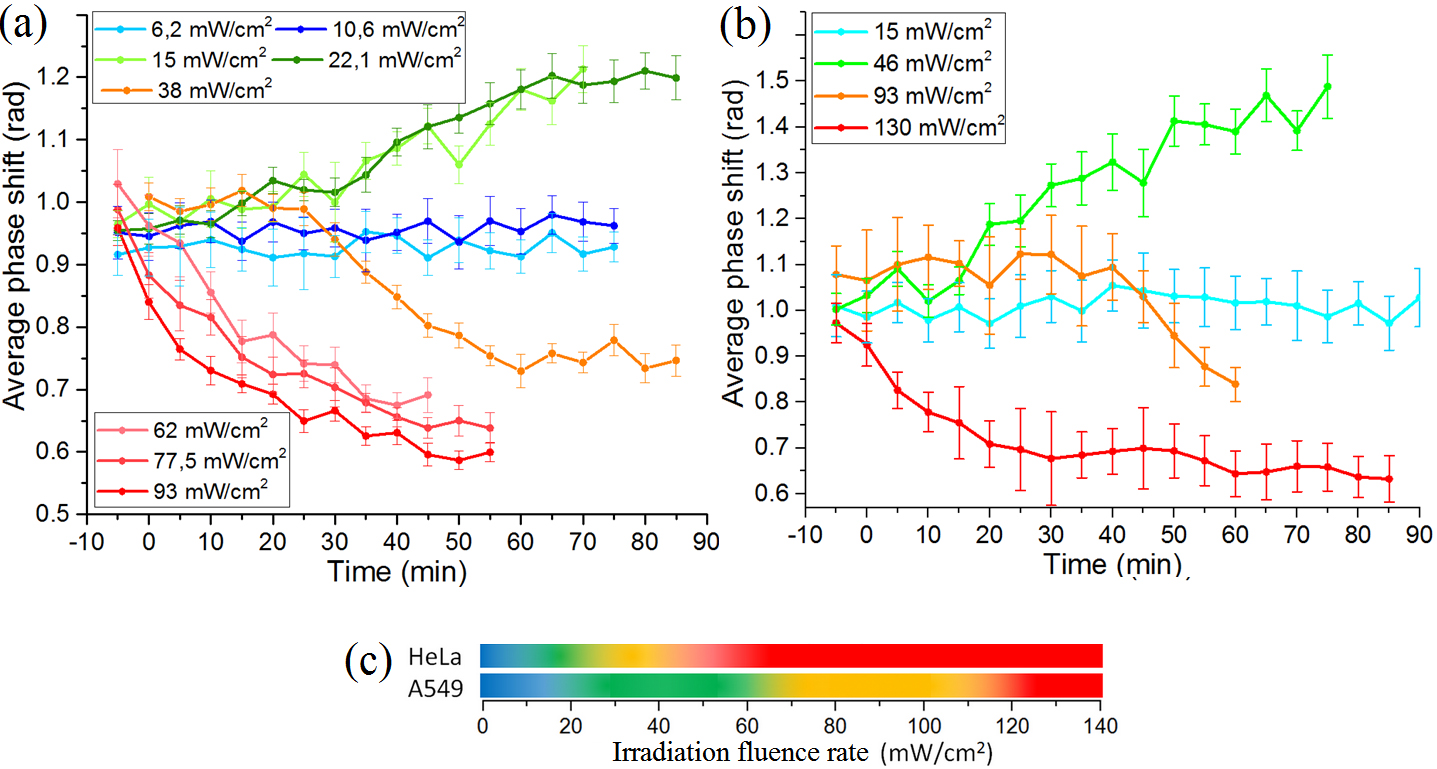}
\caption{Average phase shift dynamics in HeLa (a) and  A549 (b) cells at the indicated irradiation doses. (c) Schematics of average phase shift variation scenario as a function of fluence rate for the two cell lines. Colors in (c) correspond to those on the graphs in (a,b).}
\label{Results}
\end{figure}

The dynamics of average phase shift in living cells exposed to PD treatment at various doses was monitored during 1.5 hours. The treatment dose was varied by changing the irradiation fluence rate at the same irradiation duration (5 minutes throughout all experiments).
The observed dynamics of phase shift in the two types of cells at different irradiation doses are shown in Fig. \ref{Results}(a,b). Each experimental point on the graphs in Fig. \ref{Results}(a,b) corresponds to the value averaged over 50 cells and error bars indicate standard error.

Four substantially dissimilar scenarios of cells' response at different excitation fluence rates have been distinguished for both cell cultures.

\begin{enumerate}
\item At very low fluence rates  (6.2-10.6 mW/cm$^2$ for HeLa and 15 mW/cm$^2$ for A549 cells) no significant changes of phase shift were observed. Fluorescent images obtained with the AO/EB test assay did not reveal any significant difference in cellular morphology between control, non-treated HeLa cells (Fig. \ref{Confocal}(a)) and PS-loaded cells irradiated at these fluence rates (Fig. \ref{Confocal}(b)).

\item Low fluence rates (15-22.1 mW/cm$^2$ for HeLa and 46 mW/cm$^2$ for A549 cells) resulted in a slow increase of average phase shift up to a plateau level (1.2 rad for HeLa and 1.4 rad for A549). This process was delayed for 10-15 minutes after irradiation and was observed only in a part of the monitored cells. The phase shift increase was accompanied by the decrease of cell projected area while the cellular dry mass remained unaltered. The average value of cellular dry mass measured in HeLa cells in 70 minutes after irradiation was found to be 275 pg which is close to the initial value of 271 pg.
Typical phase images of HeLa cells taken before and after treatment at these doses  are shown in Fig. \ref{HolographyImages}(a,b). Corresponding confocal fluorescent images with AO/EB staining obtained in 60 minutes after irradiation demonstrate cell rounding and blebbing, see Fig. \ref{Confocal}(c). An absence of EB fluorescence evidences cellular membrane integrity which is in a good agreement with the cell dry mass invariance.

\begin{figure}[htbp]
\centering
\includegraphics[width=9cm]{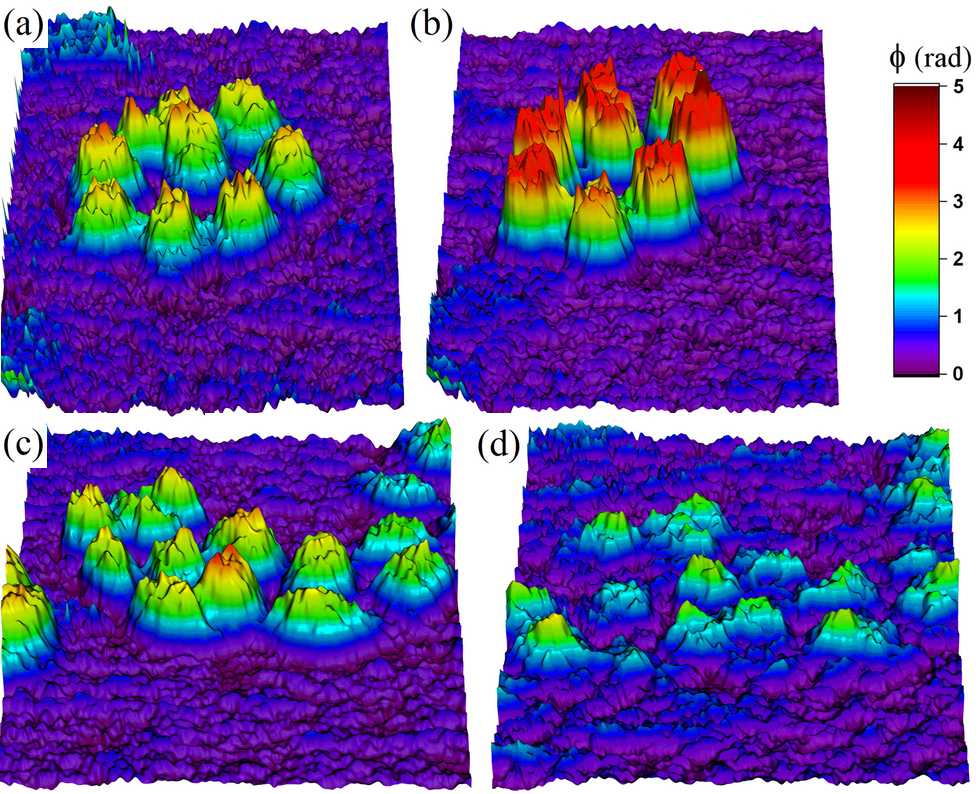}
\caption{Typical phase images of HeLa cells obtained before photodynamic treatment (a,c) and in 60 minutes after irradiation at 22.1 mW/cm$^2$ (b)  and 93 mW/cm$^2$ (d).}
\label{HolographyImages}
\end{figure}

\item Further increase of fluence rate up to 38 mW/cm$^2$ for HeLa and 93 mW/cm$^2$ for A549 cells led to a quite different trend of average phase shift evolution. The delayed decrease of phase shift was observed
starting in 25 min after irradiation in HeLa cells and in 40 min in A549. Moreover, a decrease of cellular dry mass was observed in both  kinds of cells. In HeLa cells the dry mass decrease measured at the end of the observation time amounted $ \approx 80 pg$ (30\% of the initial value).

\item At high fluence rates (62-93 mW/cm$^2$ for HeLa and 130 mW/cm$^2$ for A549 cells) a prominent decrease of the average phase shift was observed starting right after or even in the course of irradiation. The decrease in more than 0.06 rad during the first five minutes was recorded in HeLa cells. The overall  decrease of the average phase shift achieving a plateau at the level of 0.6-0.7 rad was observed in both types of cells. The observed phase shift dynamics was accompanied by a substantial reduce of cellular dry mass in $ \approx 100 pg$ (37\% of the initial value). The AO/EB fluorescent images demonstrated most of cell nuclei being stained by Etidium bromide, indicating fatal disintegration of cellular membranes, see Fig. \ref{Confocal}(d).
\end{enumerate}

Note that within the cell response scenarios 2-4 mentioned above the phase shifts recorded at the end of the observation time differed reliably from each other with the level of significance p$<$0.05.

\begin{figure}[htbp]
\centering
\includegraphics[width=8 cm]{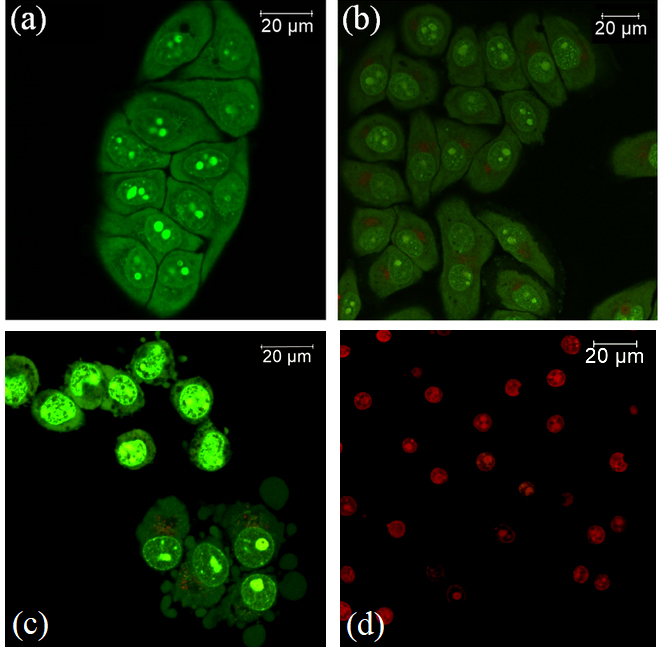}
\caption{AO/EB fluorescent images of HeLa cells before (a) and after irradiation at the fluence rate of 9.3 mW/cm$^2$ (b), 15.6 mW/cm$^2$ (c) and 100 mW/cm$^2$ (d).}
\label{Confocal}
\end{figure}

\section{Discussion}

The results obtained can be interpreted in terms of different pathways of cell response  at varying doses of PD treatment. The tendencies observed are similar for both types of cells however specific trends happen at substantially different treatment doses.

The invariance of average phase shift and cellular dry mass at very low irradiation doses (shown by cyan and blue curves in Figs.1 a and 1 b) indicates no cellular response to the treatment. This conclusion was confirmed by fluorescent images  where no cell changes are observed (see Fig. 3b for HeLa cells). This cell behavior can be explained by low amounts of generated reactive oxygen species that can be successfully deactivated by intracellular antioxidant mechanisms and by the threshold nature of cell response to PDT.

The increase of average phase shift at higher irradiation doses, shown by green curves in Figs. 1a and 1b and by phase images in Figs. 2a and 2b, along with the invariance of cellular dry mass and total phase shift can be explained by the decrease of cell projected area and cell rounding. This explanation is supported by the fluorescent image in Fig. 3c demonstrating cells rounding, blebbing and membrane integrity.
All these features allow us to suppose that apparently cells become unable to withstand the effect of generated ROS and the pathway of programmed cell death through apoptosis is activated. Note that similar variations in phase images of living cells caused by other factors were observed recently by other researchers and referred to as early apoptosis \cite{kemmler2007noninvasive,kemper2006modular}.

The dramatic decrease of the phase shift at high irradiation doses, shown by red curves in Figs. 1a and 1b and phase images in Figs. 2c and 2d, accompanied by the decrease of cellular dry mass is indicative of cell death through necrosis.
The cellular dry mass value indicates the amount of intracellular content and in normal conditions should not decrease rapidly, when averaging over several dozens of living cells.
A significant decrease of this value can be explained by cellular membrane rupture and efflux of intracellular content. The process is accompanied by the decrease of total phase shift (which is proportional to cellular dry mass) and formation of big 'blebs'. This explanation is supported by the fluorescent image in Fig. 3d clearly demonstrating cell membrane rupture.

The dynamics of average phase shift decrease can be used for estimation of intracellular content efflux rate. At high irradiation doses the severe membrane damage occurs being caused by lipid peroxidation by ROS. The intracellular protective antioxidant mechanisms are unable to recover the damage and cell dies through a fast unprogrammed mechanism, necrosis. The higher is the dose the faster happens the cell membrane rupture and the more rapid is the efflux of intracellular content. The resulting loss of dry mass however is almost the same for the specific type of cells.

Yellow curves in Figs. 1a and 1b obtained at intermediate irradiation doses showing a postponed decrease of average phase shift and loss of cellular dry mass with a pre-necrotic phase during first 20-40 minutes may be interpreted as secondary necrosis \cite{Silva2010sec-nec}.

The designated scenarios of cells response to photodynamic treatment were found to be similar for the two types of cancer cells used in our experiments. However the  irradiation doses related to the specific tendencies described above were substantially different for these cell types (see Fig. \ref{Results}(c)). A549 cells were found to be more resistant to photodynamic treatment, with all the above processes occurring at significantly higher doses than in HeLa cells.

\section{Conclusions}

We have thus performed a thorough monitoring of cells behavior under photodynamic treatment \textit{in vitro} at irradiation doses varied in a wide range. Three major pathways of cell death were considered for the two cultured cancer cell lines. Main scenarios of alterations of optical parameters of cells were identified and analyzed.
The conclusion made about changes of cellular parameters in response to treatment was based on the statistical assay with the level of significance p$<$5$\%$.
The results obtained demonstrate that different malignancies can be differently responsive to photodynamic treatment. The desired pathway of cell death requires specific treatment parameters depending upon the cell type and applied photosensitizer.

\section{Acknowledgments}

The work was supported by the Russian Science Foundation (RSF), Project 14-13-00266.
The authors acknowledge  the Common Use Center of the Institute of Cytology for opportunity to perform experiments on the confocal fluorescent microscope Leica TCS SP5.











\end{document}